\input psfig.sty

\documentclass{article}

\begin{document}



\title{HST as a reliable astrometric tool for pulsar astronomy: the  cases of  the Vela and Geminga pulsars.} 

\author{A. De Luca $^{1}$ \and R.P. Mignani $^{2}$ \and P.A. Caraveo $^{1}$}

\date{ }

\maketitle

\noindent
$^{1}$ IFC-CNR,  Via Bassini  15,   I-20133 Milan

\noindent
$^{2}$ STECF-ESO, Karl Schwarzschild  Str.2, D8574O   Garching   b.  Munchen

\vspace{10mm}

\begin{abstract}

The quest for the distance of the Vela and Geminga pulsars yielded, so far, a set of four WFPC2 images for each object. 
The availability of couples of images taken at the same period of the year, thus affected by the same parallactic 
displacement, prompted us to use Vela and Geminga as test cases to assess the reliability of HST astrometry.

\end{abstract}

\section{Introduction.} 
Optical  astrometry becomes important  when peculiar  characteristics of the pulsar emission prevent the use of radio 
techniques, such as timing or  interferometry, to measure accurate pulsar positions. This is the case for the Vela pulsar 
(PSR B0833-45), whose signal shows irregularities that affect the accuracy of the radio positioning, as well as for 
Geminga, whose radio signal appears erratic, if at all present (Kassim \& Lazio 1999 \cite{3} and references therein). 

Thus, accurate positioning of these objects has to rely on the astrometry of their faints optical counterparts. 

Over the years, the WFPC2 on HST has been repeatedly used to image both pulsars with the aim to measure their distances through 
their parallactic displacements. In order to do so, for each target a sequence of at least three observations were to be 
performed at the times of maximum parallactic elongation, 6 months apart, at the same date, with identical instrument 
set-ups. Unfortunately, obtaining three time critical observations is not always an easy task. While for the V=25.5 
Geminga, it was possible, during cycles 4 and 5, to collect the triplet of observations used by Caraveo et al. (1996)\cite{1} to 
infer the object's distance, for the V=23.6 Vela, the cycle 6 observing programme yielded only two images (collected 
in june 1997 and january 1998).

\begin{table} \begin{tabular}{c|c|c|c|c} \hline \hline
Pulsar & Obs. ID & Date & N.of exp. & Exposure(s) \\ \hline \hline
Vela & 1 & 1997 June 30 & 2 & 1300 \\
& 2 & 1998 January 2 & 2 & 1000 \\
& 3 & 1999 June 30 & 2 & 1000 \\
& 4 & 2000 January 15 & 2 & 1300 \\ \hline \hline
Geminga & 1 & 1994 March 19 & 2 & 700 \\
& & & 2 & 800 \\
& 2 & 1994 September 23 & 4 & 700 \\
& & & 2 & 800 \\
& 3 & 1995 March 18 & 4 & 700 \\
& & & 2 & 800 \\
& 4 & 1995 September 20 & 4 & 700 \\
& & & 2 & 800 \\ \hline \hline \end{tabular}

\caption{Summary of the HST/WFPC2 observations used  in the  present work.  In all cases the observations were 
taken through  filter F555W and the target was centered in the   PC chip (45 mas/px).  The columns give  the pulsar 
name, the observation ID, the   observing epoch, the total number  of  repeated exposures  for each  observation and 
the time per exposure in seconds. For each target, observations 1-3 and 2-4 define couples $
\sharp$ 1 and $\sharp$ 2, respectively}
\end{table}

The whole program has been repeated in  cycle 8 with two images available so far (June 1999 and January 2000) and a 
third one scheduled for July 2000.  The availability of two couples of WFPC2 observations of  Vela, both obtained at 
the same period  of the year, i.e.  with the  same parallactic factor, and both spanning a same time interval of 2  years 
(see Table 1), prompted us to use them to check the consistency of the procedure we have developed to superimpose 
WFPC2 images to measure proper motions and, eventually, parallaxes of isolated neutron stars. Indeed, two similar  
couples of WFPC2 observations, covering a time span of  1  year  each,  are also available  for the fainter Geminga (see 
Tab.1), for which  a fourth (unpublished) observation can be added to the triplet used by Caraveo et al.  (1996)\cite{1}. For 
both Vela and  Geminga we  can    thus obtain  two   pure,  totally independent, measurements of the proper motion, 
which can be used both  to confirm the reproducibility and consistency of WFPC2 astrometry and to assess the  weights 
of the known systematic errors possibly related to the target's brightness, the number of reference objects in the field, 
the instrument jitter, etc.

\begin{figure}
\centerline{\hbox{\psfig{figure=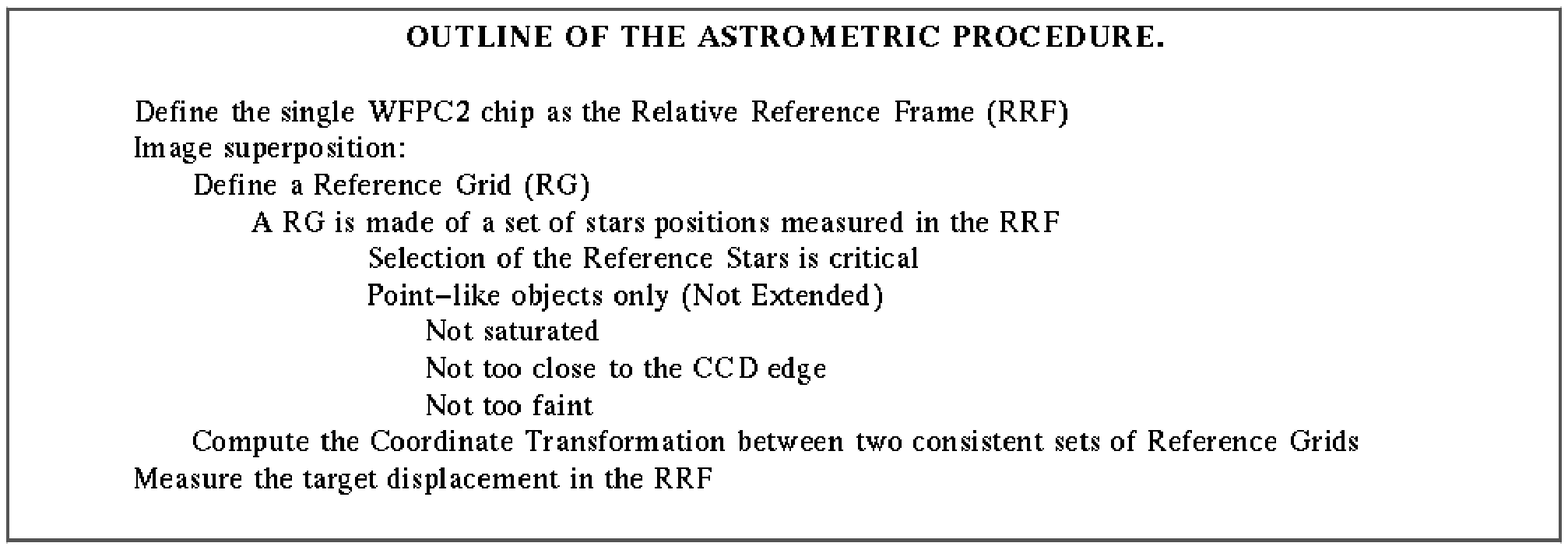,height=5.5cm,clip=}}}
  
\caption{Schematic overview of the procedure required to compute
relative astrometry measurement between sets of   images taken at different
epochs.} 
\end{figure}

\section{Data Analysis.} 
After  performing the  standard   data reduction  (OTF  recalibration, exposures association and cosmic  ray cleaning), 
we worked  on  the accurate frame superposition through a  grid  of ``good" reference stars (see Fig.1 for an outline of 
the astrometric procedure). The reference stars coordinates were computed by 2-D gaussian fitting of their intensity 
profiles,yielding a positioning accurate to within  0.01-0.07 pixels. 
\begin{table}
\small
\begin{tabular}{cccccccc} \hline \hline \\
PSR & Couple & Telescope & References & Target & $\sharp$ stars & Frame & Total \\
 & & Jitter & Centroids & Centroid & & Superposition & Accuracy \\
 & & (pixel) & Accuracy & Accuracy & & Accuracy & (pixel) \\
 & & & (pixel) &  (pixel) & & (pixel) & \\ \hline
Vela & $\sharp$1 & 0.03 & 0.01-0.06 & 0.02-0.03 & 26 & 0.04 & 0.07 \\
 & $\sharp$2 & 0.02 & 0.01-0.07 & 0.02-0.04 & 25 & 0.05 & 0.07 \\ \hline
Geminga & $\sharp$1 & 0.02 & 0.01-0.04 & 0.06-0.08 & 5 & 0.03 & 0.09 \\
 & $\sharp$2 & 0.06 & 0.01-0.04 & 0.06-0.09 & 9 & 0.08 & 0.14 \\
\hline \hline
\end{tabular}
\caption{Summary of the errors (in PC pixels) affecting the relative astrometry of the Vela pulsar and Geminga. The first 
column lists the objects names and the second the couples of observations compared (see caption of Table 1). The third 
column lists the telescope jitter, the fourth and the fifth the errors on the centroids of the reference stars and of the 
target, respectively. The sixth  reports the number of reference stars used for each couple and the seventh the error on 
the frame superposition. The overall error on the object displacement is reported in the last column.} 
\end{table}
\normalsize

Using the same procedure, we computed the coordinates of Vela and  Geminga with an average accuracy of 0.04 and 
0.08 pixels, respectively. In all cases, the overall centering accuracy depends upon several factors:  the objects S/N, the 
local background conditions, the shape of the PSF, the size of the centering area and the telescope jitter.    We then 
corrected the measured coordinates for the  effects of the instrument  geometrical distortion (Gilmozzi et al. 1995 \cite{2}) 
applying the task STSDAS/METRIC.   Next step was thus the registration of the  reference frames.  While  for Vela the 
large number  of  reference stars (25) in  the  field  (Fig. 2, left) allowed us   to follow the traditional astrometric  
approach   (fitting a  linear transformation  with  2 independent  translation factors, 2 scale  factors  for X  and Y and a 
rotation  angle), for Geminga no more than 10 stars were available (Fig. 2, right) and we reduced the number of free 
parameters by using the HST  roll angle to align the frames. The fits on the frame superposition turns out to be very 
accurate, with a rms of 0.04 and 0.08 pixels for the Vela and Geminga fields, respectively. 

In both cases the accuracy of the superposition is determined primarily by the number of reference stars and by their relative distribution, while the 
algorithm used for the coordinate transformation and  the correction for the CCD distortions play a less important role.
Table 2 summarizes the overall uncertainty on the target relative position resulting both from the combination of  the 
errors due to the centering procedure and on the frame registration.

\section{Results.}

\begin{figure}
\centerline{\hbox{\psfig{figure=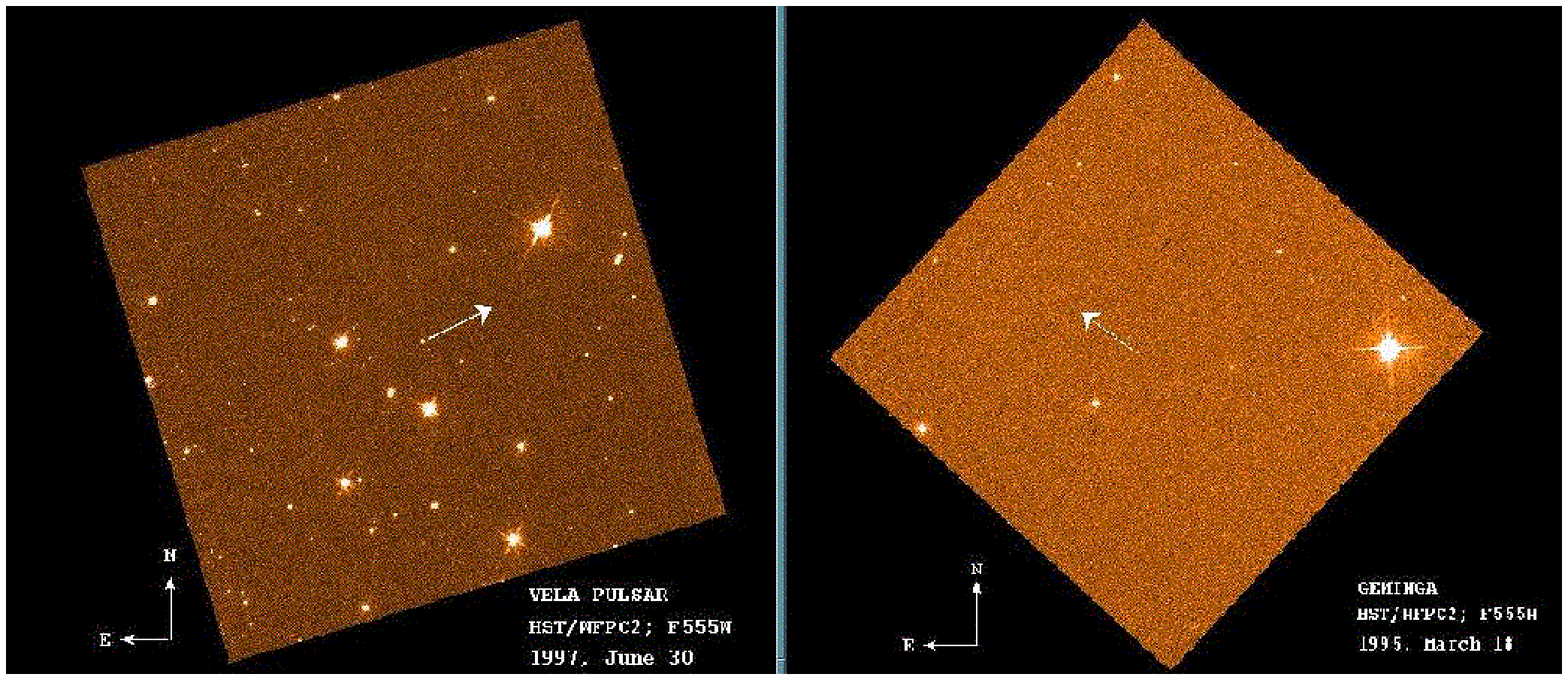,height=7cm,clip=}}}
\caption{Planetary Camera observations of the fields around the Vela and 
Geminga pulsars (left to right) taken with the 
filter 555W. The arrows marks the proper motion directions.}

\end{figure}

\begin{table} 
\begin{tabular}{cccccc} \hline \hline \\ 
PSR & Couple & PM(RA) & PM(DEC) & PM & PA \\ 
 & & mas/yr & mas/yr & mas/yr & degrees \\ \hline 
Vela & $\sharp$1 & $-45.2\pm2$ & $26.2\pm2$ & $52.2\pm3$ & \\ 
 & $\sharp$2 & $-45.6\pm2$ & $26.3\pm2$ & $52.6\pm3$ & \\ 
 & \bf{average} & \bf{$-45.4\pm1.5$} & \bf{$26.2\pm1.5$} & \bf{$52.4\pm2$} & \bf{$300\pm2$} \\ \hline 
Geminga & $\sharp$1 & $141.3\pm4$ & $94.0\pm4$ & $169.7\pm6$ &  \\ 
 & $\sharp$2 & $140.9\pm6$ & $99.1\pm6$ & $172.3\pm8$ & \\ 
 & \bf{average} & \bf{$141.1\pm4$} & \bf{$96.5\pm4$} & \bf{$171.0\pm6$} & \bf{$56\pm2$} \\ \hline \hline 
\end{tabular} 
\caption{Results. For  both Vela  and   Geminga the proper motions  values (in mas/yr) are computed independently   
using the couples of observations 1 and 2 i.e. the ones obtained in the same   period of  the year (see caption to table  
1). The   average results  with the corresponding proper motion position  angles (P.A.) in  degrees  are also reported.}  
\end{table} 
 
The Vela and Geminga displacements measured from each couple of images are reported in Table  3. The quoted 
uncertainties  are conservative and result from a standard propagation of the errors  affecting objects centroids and 
frame superpositions. In the case of Geminga, the higher uncertainties are due  mainly to the paucity of   reference 
stars, to the faintness  of the pulsar optical   counterpart and to the higher jitter value of the fouth image. Last but not 
least, the Geminga errors are diluted over a shorter time span (1 year vs. 2 years   for Vela).  All this may contribute to 
the small   difference   (5 mas/yr)  between values of $\mu_{\delta}$, measured for the two couples of images.
In  any case, the reproducibility of the results, always well within the quoted errors, gives a  further demonstration  of 
the reliability of the WFPC2 as an outstanding astrometric tool for faint targets. 

We can thus confidently compute a standard average for each couple of  measurements. The final results, also reported in Table 3, are bound to become the 
reference values for the proper motions of these two nearby isolated neutron stars.

\end{document}